\documentstyle[12pt]{article}
\begin{document}
\title{Forward diffraction dissociation of the virtual photons at small
$x_{Bj}$ in the QCD dipole picture}
\author{ A.Bialas\thanks{e-mail:bialas@thp1.if.uj.edu.pl} and
 W.Czyz\thanks{e-mail:czyz@thp1.if.uj.edu.pl} \\
M.Smoluchowski Institute of Physics\\
Jagellonian University\thanks{Address: 30-059 Krakow, Reymonta 4}}
\maketitle
\begin{abstract}
Forward differential cross-section for diffraction dissociation of virtual
photons on nucleon target is calculated in the QCD dipole picture. The
numerical estimates are presented and discussed.
\end{abstract}
\section{Introduction}

 It was shown recently \cite{na1,na2} that  the QCD dipole
picture
\cite{mu1,mu2,mu3} can successfully describe
the HERA data \cite{he1} on proton structure function $F_2$ at small $x_{\rm Bj}$
in a wide range of $Q^2$. This important observation,  indicating that the
BFKL dynamics \cite{li1} may be  relevant already in the kinematic region
accessible at the existing machine, invites one to apply the same methods for
description of other related processes. The problem is particularly interesting
in view of the conflicting opinions about the subject \cite{al1}.

Diffraction dissociation of the virtual photons,  also measured
recently at HERA \cite{he2}, presents  a challenging example of such a process.
Indeed, if one takes the traditional point of view \cite{go1} that both elastic
and inelastic diffractive amplitudes at high energies are consequences of
absorption of the incident particle waves, it follows that the total
cross-section (related to forward elastic amplitude by the optical theorem) and
the forward diffraction cross-section should be very closely related. In the
present paper we explore this relation in the framework of the QCD dipole
picture.

There are several reasons for
a particular interest in studying of the {\it forward} diffraction
dissociation in the QCD dipole picture.
\begin{description}
\item{---} First,
as we show below, the theoretical calculation does not introduce new
parameters as compared to those already known from the fit to the total
cross-section
measurements \cite{na1,na2}. Some minor ambiguities which appear in the final
result are related to the limits of the validity of the model itself and thus
may serve as important clues for the extension of the whole picture.
\item
{---} Second, the relation between the total cross-section and diffraction
dissociation
 being quadratic, their comparison provides a sensitive test of the
model, in particular of the absolute normalization of the cross-sections,
difficult to obtain by any other method. In view of the possibility that
measurements of the forward diffraction dissociation may be soon available
\cite{es1}, our exercise can also be of some practical importance.
\item
{---} Third, the BFKL dynamics implies that the behaviour of the forward
amplitudes cannot be obtained by a simple extrapolation of the high-energy
limit found  at
finite momentum transfer \cite{mu2,ba1}. It is thus interesting to investigate
separately the point $p_t=0$.
\item
{---} Fourth, forward diffraction dissociation is the process which dominates the
shadowing of virtual photons in nuclei. It is therefore important to know it as
well as possible if one wants to study the multiple scattering effects.
\item
{---} Finally, the calculation at $p_t=0$ is technically simpler than that in the
general case of arbitrary momentum transfer. This permits to have a much better
control
of the underlying assumptions and of the restrictions implied by the limited
validity of the QCD dipole picture.
\end{description}

Recently, using the QCD dipole picture, we have derived the formulae for
nuclear shadowing of high-energy virtual photons \cite{bi1}. As noted
above, this process is
closely related to the diffractive phenomena.
We shall explore this relation in the present paper.

Diffraction dissociation in the framework of the BFKL dynamics was already
discussed extensively  \cite{ba1,ba2} using the
momentum representation of the amplitudes. Our approach, based
on the QCD dipole technique \cite{mu1,mu2,mu3}, is closer to that of
\cite{ni1,ge1} where the impact parameter representation is used instead.

In the next section we explain the meaning of the two components of diffractive
dissociation, as defined in \cite{bi2,bi3}. Section 3 and 4 present the major
result of this paper, i.e.\  the calculation of the diffractive structure
functions at $p_t=0$. Saddle point approximation is discussed in Section 5 and
numerical results in Section 6. Our conclusions are listed in the last section.
\section{Two components of the diffractive cross-section}

 As was observed in
\cite{bi2,bi3}
({\it cf.} also
\cite{ba1,ge1}), the cross-section for diffraction
dissociation in the QCD dipole model can be written as a sum of the two
components
\begin{description}
\item{\it (i)} {\it Quasi-elastic component}, arising from the elastic scattering of the
$q\bar{q}$ pair emerging from the virtual photon. This component may be
considered as a generalization of the  Vector
Dominance Model
\cite{st1, go2}: The sum over masses of vector mesons is replaced by integral over
the masses of the $q\bar{q}$ pair. Also the  transitions
between the   states with different relative transverse momenta
of $q$ and $\bar{q}$ (i.e.\  of different masses of the system) are taken
explicitly into account. Of course, such an approximation can be
reasonable only in the continuum part of the system, outside the
prominent vector meson resonances. For the explicit calculations of the
exclusive vector meson production, see \cite{a} and \cite{b}.
\end{description}

The quasi-elastic component contributes mostly to
diffractive
excitation of the masses of the order of $Q^2$, more precisely in the region
$\beta \geq .1$ where, as usual,
\begin{equation}
\beta = \frac{Q^2}{Q^2+M^2}\,,    \label{1}
\end{equation}
and $M$ is the mass of the diffractively produced system.
\begin{description}
\item{\it (ii)} {\it Direct component}, corresponding to the direct
(two gluon exchange) interaction of the
full dipole cascade  developed from the primary fluctuation of the virtual
photon. This component can be regarded as a realization of the so-called
triple-pomeron coupling \cite{ka1} in the QCD dipole picture \cite{mu2}. It
 contributes mainly to the region of very large masses, $\beta \ll 1$.
\end{description}

\section{Quasi-elastic component}

As is clear from its definition, the amplitude for the quasi-elastic 
scattering of a virtual photon on a dipole of the transverse diameter
$r_0$, leading to a final state consisting of quark and antiquark of the
total mass $M$ and the relative transverse momentum $\vec k$,
can be  written as
\begin{eqnarray}
&&\langle\vec{k},M^2\mid T^{\rm qel} \mid Q\rangle\nonumber\\ = &&\int\limits_0^1 d\eta   \int d^2r
 \langle\vec{k},M^2\mid \vec{r},\eta\rangle
\langle\vec{r},\eta\mid T\mid \vec{r},\eta\rangle
{\mit\Psi}(\vec{r},\eta;Q)\,,
\label{2}
\end{eqnarray}
where     ${\mit\Psi}(\vec{r},\eta;Q)$ are the light-cone photon wave functions
\cite{bj1}, with $\vec r$ being the quark-antiquark relative transverse
distance and $\eta$ the light-cone momentum function of one of the
quarks,
\begin{equation}
\langle\vec{r},\eta\mid T\mid \vec{r},\eta\rangle \equiv T(r,r_0;Y) =
\pi\alpha^2  r_0^2\int\frac{d\gamma}{2\pi i} e^{\Delta(\gamma)Y}
\left(\frac{r}{r_0}\right)^{\gamma}
h(\gamma)  \label{3}
\end{equation}
is the forward dipole-dipole elastic amplitude in the QCD dipole picture
 with \cite{mu3,bi1,bi4}
\begin{equation}
h(\gamma) = \frac{4}{\gamma^2(2-\gamma)^2}   \label{4}
\end{equation}
and
\begin{equation}
\Delta(\gamma)=\frac{\alpha
N}{\pi}\left[2\psi(1)-\psi\left(1-\frac{\gamma}2\right)-\psi\left(\frac{\gamma}2\right)\right]\,, \label{4a}
\end{equation}
$\alpha$ being the strong coupling constant. Following \cite{na1} we take
$Y=\log(c/x_{\rm Bj})$, where $c$ is a constant.

Using  the argument given in \cite{bi3}, the cross-section for
scattering on the nucleon target can be expressed in the form

\begin{eqnarray}
\frac{d\sigma^{\rm qel}(p_t=0)}{dM^2 d^2p_t}
&=&
\frac{2 N_c n_{\rm eff}^2}{4\pi^2}\int d^2k \mid \langle\vec{k},M^2\mid T^{\rm qel} \mid
Q\rangle\mid
^2
\nonumber \\
&=& \frac{2 N_c n_{\rm eff}^2}{4\pi^2}\int\limits_0^1 d\eta \eta(1-\eta)
\frac{d\phi_k}2
\mid G(r_0,\vec{k}(\hat{M}),\eta,Y;Q)\mid^2\,,
 \label{8}
\end{eqnarray}
where
\begin{equation}
G(r_0,\vec{k},\eta,Y;Q)=  \frac{1}{2\pi}  \int d^2r e^{i\vec{k}\vec{r}}
T(r,r_0,Y)
{\mit\Psi}(\vec{r},\eta;Q)\,.   \label{7}
\end{equation}
 The two-dimensional vector $\vec{k}(\hat{M})$ points in the direction of
$\vec{k}$ (described by the azimuthal angle $\phi_k$) and its length equals
\begin{equation}
\mid\vec{k}(\hat{M})\mid = \hat{M} \equiv M (\eta(1-\eta))^{\frac12}\,.
\label{9}
\end{equation}
The factor $2$ in the first equality of (\ref{8}) takes care of the sum over
different spin configurations of the $q\bar{q}$ pair
\footnote{This factor was missing in \cite{bi3}.},
$N_c$ results from the sum over colours, and $n_{\rm eff}^2$ takes care of the
average number of dipoles in a nucleon \cite{na1}. It is not difficult to
verify
that after integrating over $dM^2$, Eq.~(\ref{8}) gives the familiar formula for
the integrated diffractive cross-section \cite{go1}
\begin{eqnarray}
\frac{d\sigma^{\rm qel}}{d^2p_t}=  \frac{ 2 N_c n_{\rm eff}^2}{4\pi^2}\int\limits_0^1 d\eta
\int d^2r
\mid \langle\vec{r},\eta\mid T^{\rm qel} \mid\vec{r},\eta\rangle\mid ^2
\mid {\mit\Psi}(\vec{r},\eta;Q)\mid^2\,.
 \label{88}
\end{eqnarray}

Using the   formulae for ${\mit\Psi}(r,\eta;Q)$ \cite{bj1}
\begin{eqnarray}
{\mit\Psi}_{T,{\rm right}}(r,\eta;Q)&=& \frac{\sqrt{\alpha_{\rm em}}
e_{(f)}}{2\pi}\eta\hat{Q}e^{i\phi_r} K_1(\hat{Q}r)\,, \\     \label{9a}
{\mit\Psi}_{T,{\rm left}}(r,\eta;Q)&=& \frac{\sqrt{\alpha_{\rm em}}
e_{(f)}}{2\pi}(1-\eta)\hat{Q}e^{-i\phi_{r}} K_1(\hat{Q}r)\,, \\     \label{9b}
{\mit\Psi}_{L}(r,\eta;Q)&=& \frac{\sqrt{\alpha_{\rm em}}
e_{(f)}}{2\pi}2\eta(1-\eta)Q K_0(\hat{Q}r)\,,
\label{9c}
\end{eqnarray}
where $\alpha_{\rm em}$ is the electromagnetic coupling constant and $e_{(f)}$
 the charge of the quark of flavour f, one can integrate over the azimuthal
angle in
(\ref{7}). One obtains
\begin{eqnarray}
&&G_{T,{\rm right}}(\eta,k;Y;Q)=  \frac{\sqrt{\alpha_{\rm em}} e_{(f)}}{2\pi}\eta\hat{Q}
\int r dr  T(r,r_0;Y)K_1(\hat{Q}r)J_1(\hat{M}r)\,,\\    \label{9A}
&&G_{L}(\eta,k;Y;Q)=  \frac{\sqrt{\alpha_{\rm em}} e_{(f)}}{2\pi}2\eta(1-\eta)Q
\int r dr  T(r,r_0;Y)K_0(\hat{Q}r)J_0(\hat{M}r)\,,\nonumber\\    \label{9C}
\end{eqnarray}
$G_{T,{\rm left}}(\eta,k;Y;Q)$ is obtained from $ G_{T,{\rm right}}(\eta,k;Y;Q)$ by
substitution\break $\eta\rightarrow 1-\eta$.
The integrals over $dr$ can be found in \cite{ry1} and we finally obtain
\begin{eqnarray}
&&G_{T,{\rm right}}(\eta,M;Y;Q)=  \frac{\sqrt{\alpha_{\rm em}}
e_{(f)}\alpha^2}{2} r_0^2\eta\hat{Q}^{-1} \frac{M}{Q}
 \int\limits_{c-i\infty}^{c+i\infty} \frac{d\lambda}{2\pi i} e^{\Delta(\lambda)Y}
\nonumber \\
&&\times\left(\frac2{\hat{Q} r_0}\right)^{\lambda}
 H_T(\lambda) F\biggl(2+\frac{\lambda}2,
 1+\frac{\lambda}2;2;-\frac{M^2}{Q^2}\biggr)\,,
\label{9i}
\end{eqnarray}
\begin{eqnarray}
&&G_{L}(\eta,M;Y;Q)=
 \sqrt{\alpha_{\rm em}} e_{(f)}\alpha^2 r_0^2Q^{-1}
 \int\limits_{c-i\infty}^{c+i\infty} \frac{d\lambda}{2\pi i} e^{\Delta(\lambda)Y}
\nonumber \\
&&\times\left(\frac2{\hat{Q} r_0}\right)^{\lambda}
 H_L(\lambda) F\biggl(1+\frac{\lambda}2,
 1+\frac{\lambda}2;1;-\frac{M^2}{Q^2}\biggr)\,,
\label{9iii}
\end{eqnarray}
where
\begin{equation}
H_L(\lambda)={\mit\Gamma}^2(1+\frac{\lambda}2) h(\lambda), \;\;
H_T(\lambda)=(1+\frac{\lambda}2)H_L(\lambda)  \label{91}
\end{equation}
and $F(a,b,c,;z)$ is the hypergeometric function.

The next step is to introduce (15), (16) into (\ref{8}) and
perform the integration over $d\eta$ (integration over $d\phi_k$ gives $2\pi$
because the integrand does not depend on $\phi_k$). Using the relation
\begin{equation}
\frac{dF}{d^2p_t}= \frac{Q^2}{4\pi^2\alpha_{\rm em}} x_P^{-1}
\frac{d\sigma}{dyd^2p_t}= \frac{Q^4}{4\pi^2\alpha_{\rm em}}
x^{-1}\frac{d\sigma}{dM^2d^2p_t}    \label{8t}
\end{equation}
with $x_P=x_{\rm Bj}/\beta$,
we finally obtain for the structure functions
\begin{eqnarray}
&&\frac{dF^{\rm qel}_T}{ d^2p_t}(Q^2,x_P,\beta,p_t=0)=
\frac{ e_f^2\alpha^4 N_c}{16\pi^3}n_{\rm eff}^2  \frac{Q^{2} r_0^4}{x_P}
\nonumber \\
&&\times (1-\beta) \int\limits_{c-i\infty}^{c+i\infty} \frac{d\lambda_1}{2\pi i}
e^{\Delta(\lambda_1)Y}
 H_T(\lambda_1) \left(\frac{2\sqrt{\beta}}{Qr_0}\right)^{\lambda_1}
 F(1+\frac{\lambda_1}2, -\frac{\lambda_1}2;2;1-\beta)
\nonumber \\
&&\times \int\limits_{c-i\infty}^{c+i\infty} \frac{d\lambda_2}{2\pi i} e^{\Delta(\lambda_2)Y}
 H_T(\lambda_2) \left(\frac{2\sqrt{\beta}}{Qr_0}\right)^{\lambda_2}
F(1+\frac{\lambda_2}2, -\frac{\lambda_2}2;2;1-\beta)
K_T(\lambda)\,,\nonumber\\
 \label{8x}
\end{eqnarray}
\begin{eqnarray}
&&\frac{dF^{\rm qel}_L}{ d^2p_t}(Q^2,x_P,\beta,p_t=0)=
\frac{e_f^2\alpha^4 N_c}{8\pi^3}  n_{\rm eff}^2 \frac{Q^{2} r_0^4}{x_P} \nonumber
\\
&&\times \beta\int\limits_{c-i\infty}^{c+i\infty} \frac{d\lambda_1}{2\pi i}
e^{\Delta(\lambda_1)Y}
 H_L(\lambda_1)
\left(\frac{2\sqrt{\beta}}{Qr_0}\right)^{\lambda_1}
 F(1+\frac{\lambda_1}2, -\frac{\lambda_1}2;1;1-\beta)
\nonumber \\
&&\times \int\limits_{c-i\infty}^{c+i\infty} \frac{d\lambda_2}{2\pi i} e^{\Delta(\lambda_2)Y}
 H_L(\lambda_2)
\left(\frac{2\sqrt{\beta}}{Qr_0}\right)^{\lambda_2}
 F(1+\frac{\lambda_2}2,
 -\frac{\lambda_2}2;1;1-\beta)K_L(\lambda)\,,\nonumber\\
 \label{8y}
\end{eqnarray}
where $\lambda=\lambda_1+\lambda_2$ and
\begin{equation}
K_L(\lambda)=\frac{{\mit\Gamma}^2(2-\frac{\lambda}2)}{{\mit\Gamma}(4-\lambda)}, \;\;
K_T(\lambda)=\frac{2-\frac{\lambda}2} {1-\frac{\lambda}2}
K_L(\lambda)\,.
    \label{a8x}
\end{equation}
$e_f^2$ is the sum of squares of the quark charges.

The hypergeometric functions from (15), (16) were transformed into
those in
(19), (20) using the relation  \cite{ab1}
\begin{equation}
F(b+n,b;n+1;1-\frac{1}{\beta}) =\beta^b F(b,1-b;1+n;1-\beta)\,,   \label{8z}
\end{equation}
where $n=0$ for longitudinal and $n=1$ for the transverse component.

\section{Direct (3-Pomeron) component}

Cross-section for the direct component can be written as
\begin{eqnarray}
\frac{d\sigma^{\rm dir}}{dy d^2p_t}= n_{\rm eff}^2 \int\limits_0^1 d\eta \int d^2r \mid
{\mit\Psi}(r,\eta;Q)\mid ^2 \frac{\sigma_d(r,r_0, Y,y;p_t)}{dyd^2p_t}\,,     \label{4.1}
\end{eqnarray}
where $\sigma_d(r,r_0, Y,y)$ is the cross-section  for diffractive
dissociation of the dipole of transverse diameter $r$, scattering off the dipole
of transverse diameter $r_0$, with
$Y-y =- \log\beta$.

The formula for $\sigma_d(r,r_0, Y,y;p_t)$  at $p_t=0$ can be derived using the
Mueller--Patel formulation \cite{mu2}. We have
\begin{eqnarray}
&&\frac{\sigma_d(r,r_0, Y,y;p_t=0)}{dyd^2p_t}= \frac1{4\pi^2}\int\frac{dx_1}{x_1}
\frac{dx_2}{x_2} \frac{dx_1'}{x_1'} \frac{dx_2'}{x_2'} \nonumber \\
&&\times \tau(x_1,x_1') \tau(x_2,x_2') \rho_1(r_0,x_1,y^*) \rho_1(r_0,x_2,y^*)
\rho_2(r,x_1',x_2';Y-y^*,y-y^*)\,,\nonumber\\                       \label{in1}
\end{eqnarray}
where $\rho_1(r_0,x,y^*)$ is the single dipole density in the dipole of the
transverse diameter $r_0$ given by
\begin{equation}
\rho_1(r_0,x,y^*) =\int\limits_{c-i\infty}^{c+i\infty} \frac{d\lambda}{2\pi i}
e^{\Delta(\lambda)y^*} \left(\frac{r_0}{x}\right)^{\lambda}\,,   \label{in2}
\end{equation}
$\tau(x,x')$ is the forward scattering amplitude for collision of two dipoles
of transverse diameters $x$ and $x$' (given in \cite{mu2}), and
$\rho_2(r,x_1',x_2';Y-y^*,y-y^*)$
is the two-dipole density in an dipole of transverse diameter $r$.  The
equation for
$\rho_2(r,x_1',x_2';Y,y)$ was written down in \cite{mu2}. It reads
\begin{eqnarray}
&&\rho_2(r,x_1',x_2';Y,y) \nonumber \\
&&= \frac{\alpha N_c}{\pi^2}   {\mit\Theta}(Y-y)
\int \frac{x_{01}^2 d^2x_2}{x_{02}^2x_{12}^2} e^{-(Y-y){\mit\Omega}}
\rho_1(x_{02},x_1',y) \rho_1(x_{12},x_2',y)  \nonumber \\
&&+ \frac{\alpha N_c}{\pi^2}
\int \frac{x_{01}^2 d^2x_2}{x_{02}^2x_{12}^2}\int\limits_y^Ydy' e^{-(Y-y'){\mit\Omega}}
\rho_2(x_{02},x_1',x_2';y',y)\,,         \label{in3}
\end{eqnarray}
where ${\mit\Omega}$ is the ultraviolet cut-off defined in \cite{mu2} (the final
result does not depend on ${\mit\Omega}$). Eq.~(\ref{in3}) shows clearly the physical
meaning of $\rho_2(r,x_1',x_2';Y,y)$: the first term describes  emission
of a single gluon at rapidity $y$ followed by independent evolution of two
dipoles (formed from the produced gluon and the original $q\bar{q}$ pair), each
producing $\rho_1$ new dipoles; The second term describes the evolution of the
original dipole from rapidity $Y$ until $y$.

It is important to realize that the Eq.~(\ref{in3}) can be trusted only if
$y$ is not too close to $Y$. Indeed, the basic assumption of the BFKL
approximation is that the longitudinal momenta of emitted gluons are {\it
strongly ordered}, i.e.\  the momentum of each emitted gluon is much smaller than
that of the previous one. In the case we consider, the point $y=Y$ corresponds
to the first emitted gluon taking {\it all}  momentum of the initial quarks.
This is of course in blatant contradiction to  the very idea of the BFKL
physics. The problem can be cured by introducing an appropriate cut-off which
would eliminate such unphysical configurations. Unfortunately, there is no
unique way to do it and thus the result is ambiguous. In order to obtain an
idea about the scale of this uncertainty, we have considered a modified
equation for $\rho_2(r,x_1',x_2';Y,y)$ in which we replaced the  function
${\mit\Theta}(Y-y)$ in the
first term on the RHS of (\ref{in3}) by a  function $\omega(Y-y)$ representing
the required  cut-off  in longitudinal momentum of the emitted gluon. We have
then studied the results varying the cut-off. They are described in
Section 6.

Following closely the method shown in  \cite{bi1},
Eq.~(\ref{in3}) can be  solved with the result
\begin{eqnarray}
&&\rho_2(r,x_1',x_2';Y,y)  = \frac{\alpha N_c}{\pi^2}
\int\limits_{c-i\infty}^{c+i\infty} \frac{d\lambda}{2\pi i} r^{\lambda}
 e^{\Delta(\lambda)(Y-y)} \int\limits_0^{\infty}dx_{01}x_{01}^{-1-\lambda}\nonumber \\
&&\times\int \frac{x_{01}^2 d^2x_2}{x_{02}^2x_{12}^2}
\rho_1(x_{02},x_1',y) \rho_1(x_{12},x_2',y) \theta(\lambda,Y,y)\,,
\label{in4}
\end{eqnarray}
where
\begin{equation}
\theta(\lambda,Y,y) =\int\limits_{-y}^{Y-y} e^{-\Delta(\lambda)u}\frac{d\omega(u)}{du}
du\,.  \label{in5}
\end{equation}
For $\omega(u)\equiv {\mit\Theta}(u)$ (as in (\ref{in3})), $\theta(\lambda,Y,y)$
reduces trivially to    ${\mit\Theta}(Y-y)$.

When (\ref{in4}) is introduced into (\ref{in1}) the integrals can be explicitly
performed and the final result is \cite{bi1}
\begin{eqnarray}
&&\frac{\sigma_d(r,r_0, Y,y;p_t=0)}{dyd^2p_t}=\frac{\alpha^5 N_c}{4\pi} r_0^4
 \int\limits_{c-i\infty}^{c+i\infty} \frac{d\lambda_1}{2\pi i} e^{\Delta(\lambda_1)y}
h(\lambda_1)G(\lambda_1)
 \nonumber \\
&&\times \int\limits_{c-i\infty}^{c+i\infty} \frac{d\lambda_2}{2\pi i} e^{\Delta(\lambda_2)y}
h(\lambda_2)G(\lambda_2)
 e^{\Delta(\lambda)(Y-y)}
\left(\frac{r}{r_0}\right)^{\lambda}\frac{\theta(\lambda,Y,y)}{G(\lambda)}\,,
\label{4.2}
\end{eqnarray}
 where $\lambda= \lambda_1+\lambda_2$ and
$G(\lambda)=\frac{{\mit\Gamma}(\lambda/2)}{{\mit\Gamma}(1-\lambda/2)}$.

Averaging over the photon wave functions can also be explicitly performed and
one finally obtains
\begin{eqnarray}
\frac{d\sigma^{\rm dir}_{T,L}(p_t=0)}{dy d^2p_t}= n_{\rm eff}^2  \sigma^*_{T,L}(Q,r_0,
Y,y;p_t=0)\,,
\label{10x}
\end{eqnarray}
where $\sigma^*_{T,L}$ is given by (\ref{4.2}) with the substitution
\begin{eqnarray}
r^{\lambda}\rightarrow \frac{N_c \alpha_{\rm elm} e^2_f}{\pi}
 \left(\frac2{Q}\right)^{\lambda}
I_{T,L}(\lambda)    \label{4.3}
\end{eqnarray}
with
\begin{equation}
I_L(\lambda)=2\frac{{\mit\Gamma}^2(2-\frac{\lambda}2) {\mit\Gamma}^4(1+\frac{\lambda}2)}
{{\mit\Gamma}(2+\lambda) {\mit\Gamma}(4-\lambda)}, \;\;
I_T(\lambda)=\frac{(2-\frac{\lambda}2)(1+\frac{\lambda}2)}
{\lambda(1-\frac{\lambda}2)} I_L(\lambda)\,.
    \label{4.4}
\end{equation}
The indices $T,L$ refer to the transverse and longitudinal photons,
respectively.

Using the relation (\ref{8t}) we can thus write for the diffractive structure
function
\begin{eqnarray}
&&\frac{dF_{T,L}^{\rm dir}}{d^2p_t}(Q^2,x_P,\beta,p_t=0)=
\frac{\alpha^5 N_c^2e_f^2}{16\pi^4}n_{\rm eff}^2\frac{Q^{2} r_0^4}{x_P}
 \int\limits_{c-i\infty}^{c+i\infty} \frac{d\lambda_1}{2\pi i} e^{\Delta(\lambda_1)y}
h(\lambda_1)G(\lambda_1)
 \nonumber \\
&&\times \int\limits_{c-i\infty}^{c+i\infty} \frac{d\lambda_2}{2\pi i} e^{\Delta(\lambda_2)y}
h(\lambda_2)G(\lambda_2)
 e^{\Delta(\lambda)(Y-y)}
\left(\frac{2}{Qr_0}\right)^{\lambda}\frac{I_{T,L}(\lambda)}{G(\lambda)}
\theta(\lambda,Y,y)\,.\label{4.5}
\end{eqnarray}

\section{Saddle-point approximation}

 Before  discussing the numerical results,
it is interesting
to consider the saddle-point approximation which gives the limiting behaviour
of structure functions when
$x_P\rightarrow 0$.

For $F^{\rm qel}_L$ the calculation is straightforward, we simply evaluate
independently the two integrals. The position of  the saddle point (identical
for
$\lambda_1$ and for $\lambda_2$) is
\begin{equation}
\lambda_c= 1+ a(Y)\log\left(\frac{Qr_0}{2\sqrt{\beta}}\right) >1\,,  \label{5.1}
\end{equation}
where
\begin{equation}
a(Y)= \frac{1}{\Delta''(1)Y} =\frac{\pi}{7\alpha N_c \zeta(3)Y}
\label{5.1.1}
\end{equation}
so that we obtain
\begin{eqnarray}
&&\frac{dF_L^{\rm qel}}{d^2p_t}= \frac {e_f^2\alpha^4N_c}  {8\pi}
r_0^2n_{\rm eff}^2 c^{2\Delta_P}  x_P^{-1-2\Delta_P}\frac{2a(Y)}{\pi} \nonumber \\
&&\times\beta^{2-2\Delta_P}
F^2(3/2,-1/2;1;1-\beta)
\exp\left(-a(Y)log^2\left(\frac{Qr_0}{2\sqrt{\beta}}\right)\right)\,. \label{ii}
\end{eqnarray}
We see from (\ref{ii}) that $F_L^{\rm qel}$ shows a typical BFKL power-law
behaviour in
$x_P$
(modified by logarithmic corrections, similarly as in case of $F_2(Q^2,x_{\rm Bj})$
\cite{na1,bi5}). The
$\beta$ dependence is also dominated by the power-law
modified by a hypergeometric function. The scaling is violated by the last
factor which decreases with increasing $Q^2$ and mixes in a complicated way the
$Q-,\beta-$ and the energy dependence.

In the case of $F^{\rm qel}_T$ the simple procedure used above fails because the
integrand has
a pole at $\lambda_1+\lambda_2 =2$ and thus the contour of integration can be
moved to the position of saddle point only after the contribution from the pole
is added. To obtain this contribution we first move the contour in
$\lambda_2$ complex plane to the right of the point $2-\lambda_1$. The
contribution from the pole is then easily found to be
\begin{eqnarray}
&&\frac{dF_T^{\rm qel}}{d^2p_t}_{pole}= \frac {e_f^2\alpha^4N_c}{2\pi^3}
r_0^2n_{\rm eff}^2   x_P^{-1} \beta(1-\beta)
  \int\limits_{c-i\infty}^{c+i\infty}\frac{d\lambda}{2\pi i}
e^{2\Delta(\lambda)Y}H_T(\lambda) H_T(2-\lambda) \nonumber \\
&&\times F(1+\lambda/2,-\lambda/2;2;1-\beta)
F(2-\lambda/2,\lambda/2-1;2;1-\beta)\,.
\label{iii}
\end{eqnarray}
The integral over $d\lambda$ can now be evaluated by the saddle-point method
and we obtain
\begin{eqnarray}
&&\frac{dF_T^{\rm qel}}{d^2p_t}_{pole}= \frac{9 e_f^2\alpha^4N_c} {16 \pi}
r_0^2n_{\rm eff}^2 c^{2\Delta_P}  x_P^{-1-2\Delta_P}
\left(\frac{2a(2Y)}{\pi}\right)^{\frac12}
 \nonumber \\
&&\times\beta^{1-2\Delta_P}(1-\beta) F^2(3/2,-1/2;2;1-\beta)\,.
\label{iv}
\end{eqnarray}
One should add to this the integral to the right of the pole. This integral can
be estimated by the saddle point method because now the contours are to
the right of 1 (in both complex planes). The resulting contribution is
\begin{eqnarray}
&&\Delta\left(\frac{dF_T^{\rm qel}}{d^2p_t}\right)=- \frac{9
e_f^2\alpha^4N_c}{32\pi^2} r_0^2n_{\rm eff}^2 c^{2\Delta_P} x_P^{-1-2\Delta_P}
\beta^{1-2\Delta_P} (1-\beta)\nonumber \\
&&\times F^2(3/2,-1/2;2;1-\beta)
 \left(\frac1{\log(Qr_0/2\sqrt{\beta})}\right)
\exp\left(-a(Y)log^2\left(\frac{Qr_0}{2\sqrt{\beta}}\right)\right)\,.\nonumber\\
\label{1ii}
\end{eqnarray}
The interesting feature is that while the pole contribution scales  exactly,
the remaining integral increases with $Q^2$ (it decreases in absolute value but
has a negative sign). Consequently, $F^{\rm qel}_T$ increases with increasing
$Q^2$.
Otherwise the behaviour is qualitatively similar to that of $F^{\rm qel}_L$ except
for the ``kinematical'' factor $M^2/Q^2=(1-\beta)/\beta$ (implying that the
transverse component dominates over the longitudinal one everywhere except in
the vicinity of $\beta = 1$).

For $F_2^{\rm qel} = F_T^{\rm qel} + F_L^{\rm qel} $ we thus obtain a fairly complicated
$Q^2$ dependence reflecting $\beta$-dependent  mixing  of the transverse and
longitudinal contributions.

Direct component can be approximated by first performing the saddle point in
$\omega\equiv (\lambda_1-\lambda_2)/2$ and then in $\lambda$. We only give here
results without the cut-off, i.e.\  for $\theta(\lambda,Y,y)\equiv {\mit\Theta}(Y-y)$.

After first integration
($\omega_{\rm saddle}=0$) we obtain
\begin{eqnarray}
&&\frac{dF^{\rm dir}_{T,L}}{d^2p_t} = \frac{\alpha^5N_c^2e_f^2}{16\pi^4} n_{\rm eff}^2
\frac{Q^2r_0^4}{x_P}\frac1{2\pi}
 \int\limits_{-i\infty}^{+i\infty} \frac{d\lambda}{2\pi i}
\frac{\sqrt{\pi}}{\sqrt{\Delta''(\lambda/2)y}} e^{2\Delta(\lambda/2)y}
\nonumber \\
&&\times h^2(\frac{\lambda}2)
G^2(\frac{\lambda}2)   e^{\Delta(\lambda)(Y-y)}
\left(\frac2{Qr_0} \right)^{\lambda}
\frac{I_{T,L}(\lambda)}{G(\lambda)}\,.
\label{v}
\end{eqnarray}
The integral over $\lambda$ is somewhat more complicated.
 The saddle point equation is
\begin{equation}
\Delta'(\lambda_c/2)y + \Delta'(\lambda_c)(Y-y) =\log(Qr_0/2) .  \label{av}
\end{equation}
This can be solved in two steps. First we find $\lambda_0$
from the equation
\begin{equation}
\Delta'(\lambda_0/2)y + \Delta'(\lambda_0)(Y-y) =0\,.   \label{bv}
\end{equation}
Then we write $\lambda_c=\lambda_0+\delta$.
$\delta$ can be found from (\ref{av}):
\begin{equation}
\delta=\tilde{a}(Y-y,y) \log(Qr_0/2)\,,  \label{cv}
\end{equation}
where
\begin{equation}
\tilde{a}(Y-y,y)= \frac{1}{\Delta''(\lambda_0)(Y-y)+\frac12
\Delta''(\lambda_0/2)y}\,,   \label{ev}
\end{equation}
$\delta \rightarrow 0$ in the limit $y \gg 1, \;\; (Y-y) \gg 1$, so that the
approximation is correct.

When this is introduced into (\ref{v}) we obtain
\begin{eqnarray}
&&\frac{dF^{\rm dir}_{T,L}}{d^2p_t} = \frac{\alpha^5N_c^2e_f^2}{64\pi^4} n_{\rm eff}^2
\frac{Q^2r_0^4}{x_P}
\left(\frac{2\tilde{a}(0,y)}{\pi}\right)^{\frac12} e^{2\Delta(\lambda_0/2)y}
\nonumber \\
&&\times h^2(\frac{\lambda_0}2)
G^2(\frac{\lambda_0}2)   e^{\Delta(\lambda_0)(Y-y)}
\left(\frac2{Qr_0} \right)^{\lambda_0}
\frac{I_{T,L}(\lambda_0)}{G(\lambda_0)} \nonumber \\
&&\times \left(\frac{2\tilde{a}(Y-y,y)}{\pi}\right)^{\frac12}
\exp\left(-\frac12 \tilde{a}(Y-y,y)\log^2(Qr_0/2)\right)\,.   \label{dv}
\end{eqnarray}
Unfortunately, this requires solution of the Eq.~(\ref{bv}) and thus numerical
work.

The major new effect  seen from (\ref{dv}) is that here the $x_P$ and
$\beta$ dependence are no longer determined by the  BFKL pomeron
intercept\footnote{This effect was already discussed in
\cite{ba1,ba2}.}.
Instead, the relevant powers depend on
$\lambda_0$ which, in turn,
is a function of the ratio $y/Y$. Thus a rather complicated structure appears.
In figure 1 $\Delta(\lambda_0/2)$ and $\Delta(\lambda_0)$ which determine the
powers in $x_P$ and $\beta$ dependence are shown. One sees that they are always
larger than $\Delta_P$. Another interesting observation is that the
dominant factor in $Q$ dependence is $Q^{2-\lambda_0}$ instead of $Q$
found in $F_2(Q^2,x_{\rm Bj})$ \cite{mu1,mu2,na1}. One sees in  Fig. 1 that
$2-\lambda_0$
is substantially smaller than one and thus the effect is significant.
\begin{figure}[h]
\kern7.5cm
\caption{$x_P$ and $\beta$ dependences of the asymptotic formula for triple
pomeron contribution to $\frac{dF_2}{d^2p_t}(p_t=0)$. The ratios of powers of
$x_P$ and
$\beta$ are potted versus $\frac{y}{Y-y}$. $\lambda_0$ is the approximate
position of the saddle point. See text for details.}
\end{figure}

\section{Numerical estimates}

We have performed numerical estimates of the results presented in previous
sections using the parameters given in \cite{na1}.

 In Figure 2 the $\beta$ dependence of longitudinal and transverse
photon contributions to the quasi-elastic  component of
the structure function is shown. One sees that the
quasi-elastic  transverse
contribution largely dominates the longitudinal one, except at $\beta \approx
1$ where
the longitudinal one fills the dip.
\begin{figure}[htb]
\kern8.3cm
\caption{$\beta$- dependence of the longitudinal and transverse photon
contribution to the quasi-elastic component of the structure function
$\frac{dF^{\rm qel}}{d^2p_t}(p_t=0)$. Solid lines - no cutoff. Broken lines - a
cutoff
for the non-perturbative region of large transverse sizes of the photon is
introduced. For more details see the text.}
\end{figure}

Such a large value of  the quasi-elastic transverse component can be
traced to
 the presence of the pole at $\lambda_1+\lambda_2 =2$ in the integrand
of (\ref{8x}) which gives the main contribution to the integral. This
pole, in turn, arises from integration over the regions of
$\eta$ close to zero and one, where the cut-off on the transverse size of the
photon (as expressed by the Mc Donald function in (10), (11)) is
ineffective.
We thus see that an important  contribution to
the transverse quasi-elastic component comes from the non-perturbative
region of
large transverse sizes of the photon\footnote{Numerous authors pointed out
  this effect, see e.g.\  \cite{bj2}. The
problem was also noted
in \cite{bi2,bi3} where the integration over the photon size was restricted to
a finite region $r\leq r_0$.}.
To investigate this problem in   more detail, we have
calculated the structure functions under the condition
that the cutoff in photon size should never be larger than a certain parameter
R, i.e.\ 
\begin{equation}
\eta(1-\eta)Q^2 \geq \frac1{R^2}\,.   \label{41}
\end{equation}
The resulting formulae are identical to Eqs.~(19), (20) and (33)
where now
\vskip-0.5cm
\begin{eqnarray}
K_L(\lambda)\rightarrow \tilde{K}_L(\lambda)&\equiv&
\int\limits_{\epsilon}^{1-\epsilon} d\eta
[\eta(1-\eta)]^{1-\frac{\lambda}2}\,,\nonumber\\
K_T\lambda)\rightarrow \tilde{K}_T(\lambda)&\equiv&
\int\limits_{\epsilon}^{1-\epsilon} d\eta
\eta^2[\eta(1-\eta)]^{-\frac{\lambda}2}\,,\nonumber\\
I_L(\lambda)\rightarrow \tilde{I}_L(\lambda)&\equiv&
2\frac{{\mit\Gamma}^4(1+\frac{\lambda}2)} {{\mit\Gamma}(2+\lambda)}
\tilde{K}_L(\lambda)\,,
\nonumber\\
I_T(\lambda)\rightarrow \tilde{I}_T(\lambda)&\equiv&
2\frac{2+\lambda}{\lambda}\frac{{\mit\Gamma}^4(1+\frac{\lambda}2)}
{{\mit\Gamma}(2+\lambda)} \tilde{K}_T(\lambda)\,,
\label{42}
\end{eqnarray}
where
\begin{equation}
\epsilon=\frac12\left(1- \sqrt{1-\frac4{R^2Q^2}}\right)\approx
\frac1{R^2Q^2}\,.
\label{43}
\end{equation}
\vskip-0.2cm
The sensitivity to this ``non-perturbative'' contribution can be seen from
 Figure 2 where  the $\beta$ dependence following from (\ref{41},\ref{42})
for $R=r_0$ is also plotted. One sees that $F_T^{\rm qel}$ is  reduced by a
factor of about 2, in comparison
with the uncut result, whereas the longitudinal component is hardly affected.
We thus
conclude that the QCD dipole picture does not provide a reliable prediction for
$F_T^{\rm qel}$: it must be supplemented by an additional non-perturbative input.
Inverting this argument, one may say that measurements of this component can
give interesting information on non-perturbative structure in the small
$x_{\rm Bj}$ region.

The $Q$-dependence of the quasi-elastic component is shown in Figs. 3a and 3b.
One sees that indeed, as was seen qualitatively from the saddle-point
approximation, the longitudinal component decreases
 whereas the transverse one increases with increasing $Q$. In result
$F_2^{\rm qel}$ decreases at $\beta =1$, is almost constant at $\beta \approx 0.8$,
 and increases at smaller $\beta$. The comparison between  figures 4a and
4b shows that this qualitative behaviour
is unaffected by the cut (\ref{41}), although the precise rate of change of
course  depends on it.

We have also calculated the $x_P$ dependence. It shows a simple power law
behaviour,
consistent with the expectations from the saddle-point formulae of Section 5.
\begin{figure}
\kern7cm
\caption{The $Q$- dependence of the quasi-elastic component of
$\frac{dF^{\rm qel}}{d^2p_t}(p_t=0)$ for various $\beta$'s.}
\end{figure}
\begin{figure}
\kern8.5cm
\caption{The triple-pomeron contribution to $\frac{dF_2}{d^2p_t}(p_t=0)$ for various
cutoff's given by Eq.~(\ref{6n}) and discussed in  the text.}
\end{figure}

The $\beta$-dependence of the 3-Pomeron component is plotted in figure 4 in the
log-log scale. As described in Section 4, we have investigated the
dependence of the results on the cutoff in the rapidity of the emitted gluons.
\noindent The cutoff function was taken in the form
\begin{equation}
\omega(Y-y) =\left(1-e^{y-Y}\right)^k \,,    \label{6n}
\end{equation}
where $k$ is a constant.

The results in Fig.4 are shown for $k=0$ (no cut) , $k=1$ and $k=2$. One sees
that the uncut distribution follows very closely a power law, $\beta^{-.34}$,
in
the full region of $\beta$. This power is significantly larger than $\Delta_P$,
as expected from the discussion in Section 5. The effects of the cutoff
(\ref{6n}) is twofold.
First, as expected, it reduces  strongly the structure function at large
$\beta$  (at
$\beta=1$ the structure function vanishes) and thus changes its shape. The
structure function is  affected, however, also in the region of small $\beta$,
$\beta\leq .1$,  where it is simply multiplied by a constant
(smaller than one).

We thus see that without a better understanding of the energy-momentum
conservation constraints in gluon emission (which are responsible for the
cutoff in the gluon rapidity spectrum) there is little hope to obtain accurate
predictions for the 3-pomeron component of the structure function, even in the
region of small $\beta$.

Recently, Andersson \cite{an2} suggested that the energy-momentum conservation
constraints in gluon emission can be taken into account by simply cutting
out $\frac{11}6$ from the edge of the available phase-space in
rapidity.\footnote{This observation is consistent with the measurements of
rapidity distribution of particles produced in high-energy hadronic
collisions and of ``stopping power'' in hadron-nucleus collisions
\cite{bu1}.} The result of this prescription
(which amounts to take
$\omega(Y-y)\equiv
{\mit\Theta}(Y-y-\frac{11}6)$ and thus simply changes $\beta$ into $\beta
e^{-11/6}\approx .16\beta$)
is also shown in Fig.4. One sees that the value of the 3-pomeron component in
the small $\beta$ region is reduced by a factor of about $2$.

\section{Conclusions and outlook}

Using the QCD dipole picture, we have derived the formulae for  the forward
cross-section of diffraction dissociation
of a virtual photon on a dipole of transverse size $r_0$. Using the hypothesis
formulated in \cite{na1} (i.e.\  assuming that the nucleon is a collection of a
number ($n_{\rm eff}$) of QCD dipoles)  we
estimated the relevant cross-sections  on the nucleon target.
Our conclusions can be formulated as follows.
\begin{description}
\item{(a)} Measurements of the forward differential diffractive cross-section shall
provide a significant test of the applicability of the Good-Walker
 idea \cite{go1}
on the origin of diffraction dissociation.
\item{(b)} Fairly precise predictions can be made for the longitudinal quasi-elastic
  contribution. The results for other components are somewhat ambiguous. The
transverse quasi-elastic
component is sensitive to the non-perturbative region of large transverse sizes
in the
photon wave function (corresponding to the ``aligned jets'' of Ref.\cite{bj3})
and therefore depends on additional non-perturbative parameters. Its
measurements may thus provide an interesting information on non-perturbative
physics at small $x_{\rm Bj}$. There is thus a strong interest in separate
measurements of longitudinal and transverse components of the diffractive
structure functions.
The 3-Pomeron component is sensitive to the probability of gluon emission at
large rapidity and thus
to the effects of energy-momentum
conservation which are neglected in the QCD dipole model (and in the whole BFKL
approach).
\item{(c)} Within the ambiguities listed above, the obtained structure functions
 turn out to
be of the same order as  those obtained from a simple extrapolation of
the existing data \cite{he3} to $p_t=0$. The detailed comparison with the
future precise data shall thus be very illuminating.
\item{(d)} The obtained effective pomeron intercept (including logarithmic
corrections) turns out to be close
to that measured in
$F_2$, i.e., in the total $\gamma^*$ cross-section
(and even somewhat larger in the 3-Pomeron region). Thus we predict a
substantially stronger $x_P$ dependence of diffractive structure functions at
$p_t=0$ than that observed till now at larger $p_t$ \cite{he2,bi2,bi3,bi5}.
\item{(e)} Our calculation predicts a rich structure of the scaling violation.
Longitudinal quasielastic component is expected to decrease with increasing
$Q^2$. The transverse one increases to  a limiting value at large $Q^2$.
 The
3-Pomeron contribution increases,  following approximately a power law in
$Q^2$.
\end{description}

 To summarize, the QCD dipole picture predicts some  interesting
 physics to be found  in the diffractive structure functions at $p_t=0$. It
remains to be seen if this shall indeed show up in the future measurements.

\vspace{.3cm}
Several comments are in order.
\begin{description}
\item{\it(i)} The {\it forward} elastic dipole-dipole amplitude given  by (\ref{3}) is
the integral of the amplitude {\it at a fixed impact parameter}:
\begin{equation}
T(r; p_t=0)= \int d^2 b T(r;b)\,.    \label{51}
\end{equation}
The numerical MC estimates  reported in \cite{sa1} indicate that
$T(r;b)$
violates unitarity at small impact parameters. This was shown in \cite{sa1}
not to be essential for the value of the integrated amplitude $T(r;p_t=0)$
given in (\ref{51}), at least at the presently available energies. Therefore we
expect that also in our
case the possible violation of unitarity should not invalidate the obtained
estimates.
\item{\it(ii)} It was shown in \cite{mu2} that the diffractive cross-section
 calculated
at fixed $p_t$ in the 3-Pomeron region, when extrapolated to $p_t=0$, reveals a
singularity of the form
$\sim p_t^{-1}$. Our calculation shows explicitly how this singularity is
regularized by the  energy-dependent cut-off in impact-parameter
space \cite{mu3}
\begin{equation}
\log^2{b^2/r_0^2} \leq 2 [a(x_{\rm Bj})]^{-1} \equiv \frac{14\alpha N_c
\zeta(3)\log(1/x_{\rm Bj})}{\pi}\,.       \label{52}
\end{equation}
\item{\it(iii)} The leading logarithmic approximation which is at the origin of the BFKL
approach (and thus also of the QCD dipole picture) does not allow to fix
uniquely the energy scale. Therefore  relation of the variables
$x_{\rm Bj}$ and $x_P$  to $Y$ and $y$ is somewhat arbitrary. This is particularly
controversial in the case of the quasi-elastic component, where the dipole
picture suggests $Y$ as the relevant variable,
 whereas the
experience from Regge phenomenology suggests $y$ instead \cite{bi3}. We have
taken $Y$ to be fully consistent in our derivation of Eq.~(\ref{8}) (if $y$ is
taken, (\ref{88}) is inconsistent with (\ref{8})).
\item{\it(iv)} It was suggested in \cite{bi4} that rapidity variable $Y$ can be affected
by the value of the light-cone momentum fraction $\eta$ in the photon wave
function, e.g.,
\begin{equation}
Y=\log\left(\frac{c\eta(1-\eta)}{x_{\rm Bj}}\right)\,.   \label{7.1}
\end{equation}
 In the present paper we have taken the conservative approach
and neglected this possibility. It is interesting to note, however, that this
effect would strongly reduce the contribution from the regions $\eta\approx 0$
and $\eta \approx 1$. Consequently, the sensitivity of the calculations to the
non-perturbative contributions would be substantially reduced. 
\end{description}
\bigskip\medskip
      We would like to thank D. Kisielewska, A. Eskreys and R. Peschanski for
discussions.


\end{document}